\documentclass[12pt,onecolumn]{article}

\usepackage{arxiv}

\usepackage[backend=biber, style=numeric, sorting=none, sortcites=true, doi=true, url=false,giveninits=true, maxbibnames=3]{biblatex}
\usepackage[english]{babel}     
\usepackage[utf8]{inputenc}     
\usepackage[T1]{fontenc}        
\usepackage{lmodern}            
\usepackage{dcolumn}            
\usepackage{bm}                 
\usepackage{etoolbox}           
\usepackage{xcolor}             
\usepackage{enumitem}           
\usepackage[normalem]{ulem}     
\usepackage{lipsum}             
\usepackage{csquotes}
\usepackage[scr=rsfs]{mathalfa} 
\usepackage{amsmath}            
\usepackage{amsfonts}           
\usepackage{amssymb}            
\usepackage{amsthm}             
\usepackage{slashed}            
\usepackage{mathtools}          
\usepackage{ragged2e}

\usepackage{graphicx} 
\usepackage{subcaption}
\usepackage{hyperref}
\usepackage{comment}

\usepackage{cleveref}
\crefname{equation}{}{}
\Crefname{equation}{}{}
\crefformat{equation}{(#2#1#3)}
\crefrangeformat{equation}{(#3#1--#2#4)}
\crefformat{figure}{#2#1#3}
\crefmultiformat{figure}{#2#1#3}{, #2#1#3}{, #2#1#3}{ e #2#1#3}
\crefformat{table}{#2#1#3}
\crefmultiformat{table}{#2#1#3}{, #2#1#3}{, #2#1#3}{ e #2#1#3}

\makeatletter
\newcommand{\mybf}[1]{%
  \begingroup
  \ifmmode 
    \def\@tempa{#1}%
    \edef\@tempa{\noexpand\in@{#1}{abcdefghijklmnopqrstuvwxyzABCDEFGHIJKLMNOPQRSTUVWXYZ0123456789}}%
    \@tempa
    \ifin@
      \mathbf{#1}
    \else
      \pmb{#1}
    \fi
  \else
    \textbf{#1}
  \fi
  \endgroup
}
\makeatother

\makeatletter
\newif\if@insideout
\@insideoutfalse

\DeclareRobustCommand{\mysout}[1]{%
  \begingroup
  \@insideouttrue
  \ifmmode
    \expandafter\cancel
  \else
    \expandafter\sout
  \fi{#1}%
  \@insideoutfalse
  \endgroup
}
\makeatother

\newcommand{\partialt}[1]{\frac{\partial #1}{\partial t}}

\usepackage{authblk}
\title{\huge A numerical study on plasma acceleration processes with ion dynamics at the sub-nanosecond timescale}
\author[1,2,*]{G. Parise}
\author[3]{A. Cianchi}
\author[1]{M. Galletti}
\author[3]{F. Guglietta}
\author[1]{R. Pompili}
\author[4]{A.R. Rossi}
\author[3]{M. Sbragaglia}
\author[3]{D. Simeoni}

\affil[1]{INFN, Laboratori Nazionali di Frascati, Via Enrico Fermi 54, 00044, Frascati, Italy}
\affil[2]{Department of Physics, Tor Vergata University of Rome, Via della Ricerca Scientifica 1, 00133, Rome, Italy}
\affil[3]{Department of Physics \& INFN, Tor Vergata University of Rome, Via della Ricerca Scientifica 1, 00133, Rome, Italy}
\affil[4]{INFN, Section of Milan, via Celoria 16, 20133, Milan, Italy}
\affil[*]{\texttt{gianmarco.parise@lnf.infn.it}}
\date{\today}

\begin{document}

\maketitle

\begin{abstract}
Plasma wakefield acceleration is a groundbreaking technique for accelerating particles, capable of sustaining gigavolt-per-meter accelerating fields. Understanding the physical mechanisms governing the recovery of plasma accelerating properties over time is essential for successfully achieving high-repetition-rate plasma acceleration, a key requirement for applicability in both research and commercial settings.
In this paper, we present numerical simulations of the early-stage plasma evolution based on the parameters of the SPARC\_LAB hydrogen plasma recovery time experiment (Pompili {\it et al.}, \textit{Comm. Phys.} {\bf 7}, 241 (2024)), employing spatially resolved Particle-in-Cell and fluid models. The experiment reports on a non-monotonic dependence of the plasma recovery time on the initial plasma density, an effect for which ion motion has been invoked as a contributing factor. The simulations presented here provide further insight into the role of ion dynamics in shaping this behavior. Furthermore, comparing Particle-in-Cell and fluid approaches allows us to assess the quality of fluid models for describing this class of plasma dynamics.
\end{abstract}

\section{Introduction \label{sec:intro}} 
Particle accelerators have been a pivotal technology for advancing scientific research across various fields over the past century, and their use has also spread widely in industry and medicine~\cite{sheehy-2024, doyle-2019}. In this context, radio frequency technology has achieved a high level of brilliance and luminosity for accelerated beams~\cite{Ackermann2007fel,Decking2020xfel}. Nevertheless, the accelerating gradient is still limited to around $\le 100~\mbox{MV/m}$ at room temperature~\cite{Argyropoulos2018xband} because of electrical breakdown, so that reaching high energies requires huge footprints. Plasma wakefield acceleration has demonstrated the possibility to accelerate high-quality bunches to $\mathcal{O}($\mbox{GV/m}$)$ accelerating gradients~\cite{Litos20169GeV, pompili2021enspreadmin, pompili2022fel} through the wake generated by intense particle bunches or laser beams, demonstrating its potential as a groundbreaking technique. 

In order to meet the bunch requirements, for example, in terms of luminosity or brilliance, a high repetition rate is essential for plasma wakefield acceleration~\cite{hooker2013RepetitionRateNeeds}, essential in current facilities and next generation light sources like EuPRAXIA~\cite{pap21}.
Unlike radio frequency cavities, plasma wakefields are rapidly damped by nonlinear and thermal effects after only a few oscillations~\cite{DArcy2022, Gilljohann2019ionmotion, simeoni2024thermalfluid}. As a result, high-gradient and high-quality acceleration can only be achieved within the first wake. Consequently, after each acceleration stage, it is necessary to wait for the perturbed plasma to recover its original configuration before a subsequent bunch can experience comparable accelerating conditions. This defines a limit on the repetition rate capability of plasma acceleration. 

Recent experiments with argon plasma reported a recovery time of $60~\mbox{ns}$~\cite{DArcy2022}.
Ion motion has been identified as a key physical mechanism governing this recovery process. Depending on the nature of the perturbing driver, two closely related ion-dynamics scenarios are commonly distinguished in the literature: for electron bunch drivers, ions are radially drawn towards the axis by the electromagnetic fields associated with the driver itself and the ponderomotive force of the excited electron plasma wave~\cite{DArcy2022, khudiakov-2022}, an effect that has been experimentally studied using shadowgraphy techniques~\cite{Gilljohann2019ionmotion, zgadzaj-2020}; for laser-driven perturbations, ion motion towards the axis is instead triggered only by the ponderomotive force of the electron plasma wave~\cite{gorbunov2001plasma, gorbunov2003ionmotion}.
Analytical models have been developed to describe ion dynamics in the narrow driver bunch limit~\cite{vieira2014ion} as well as in regimes involving intense energy sources, including both laser and particle bunch drivers~\cite{sahai2017ionmotion}. 

\begin{figure*}[t!]
    \centering
\raisebox{0.65cm}{\includegraphics[width=0.52\textwidth]{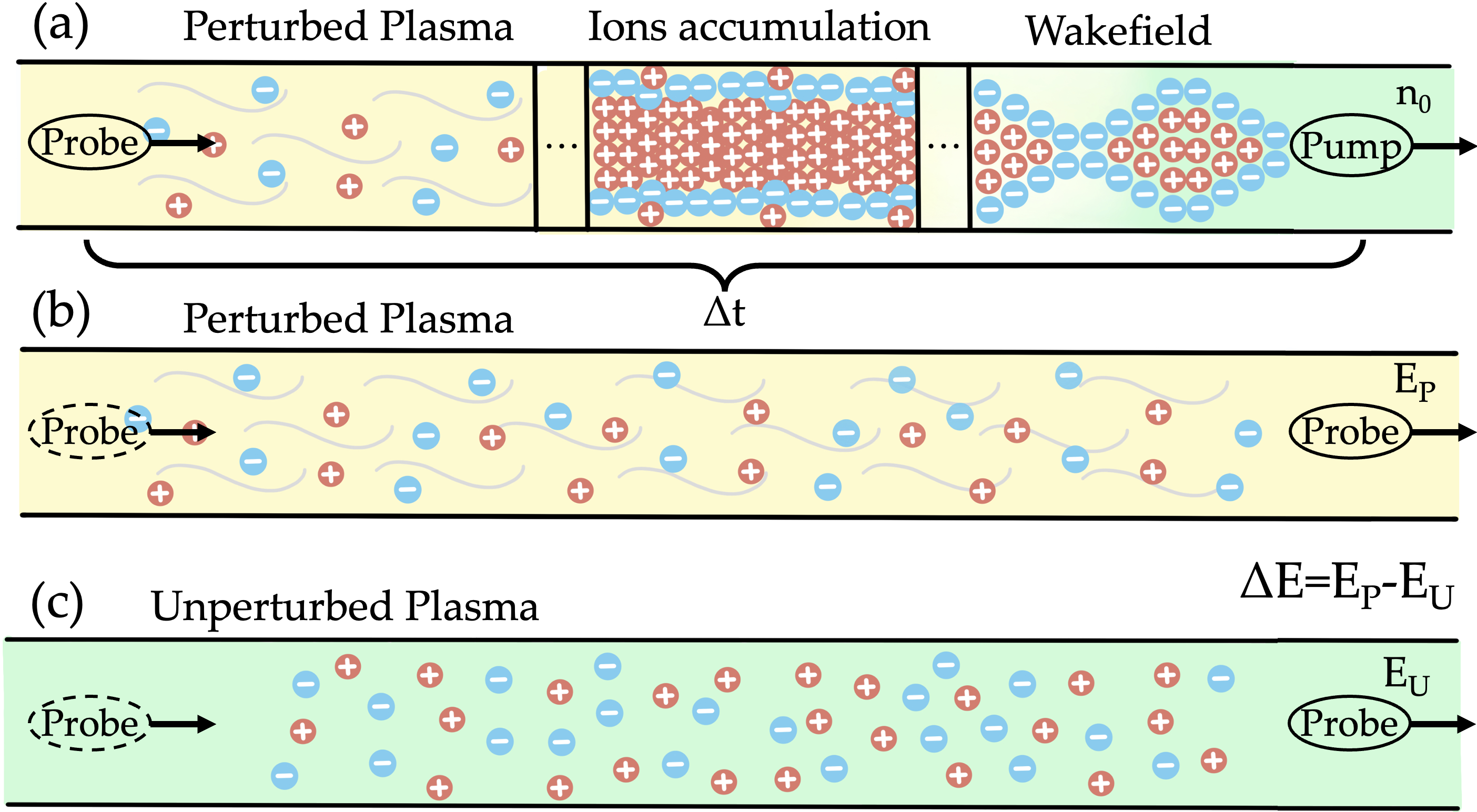}}
\includegraphics[width=0.45\textwidth]{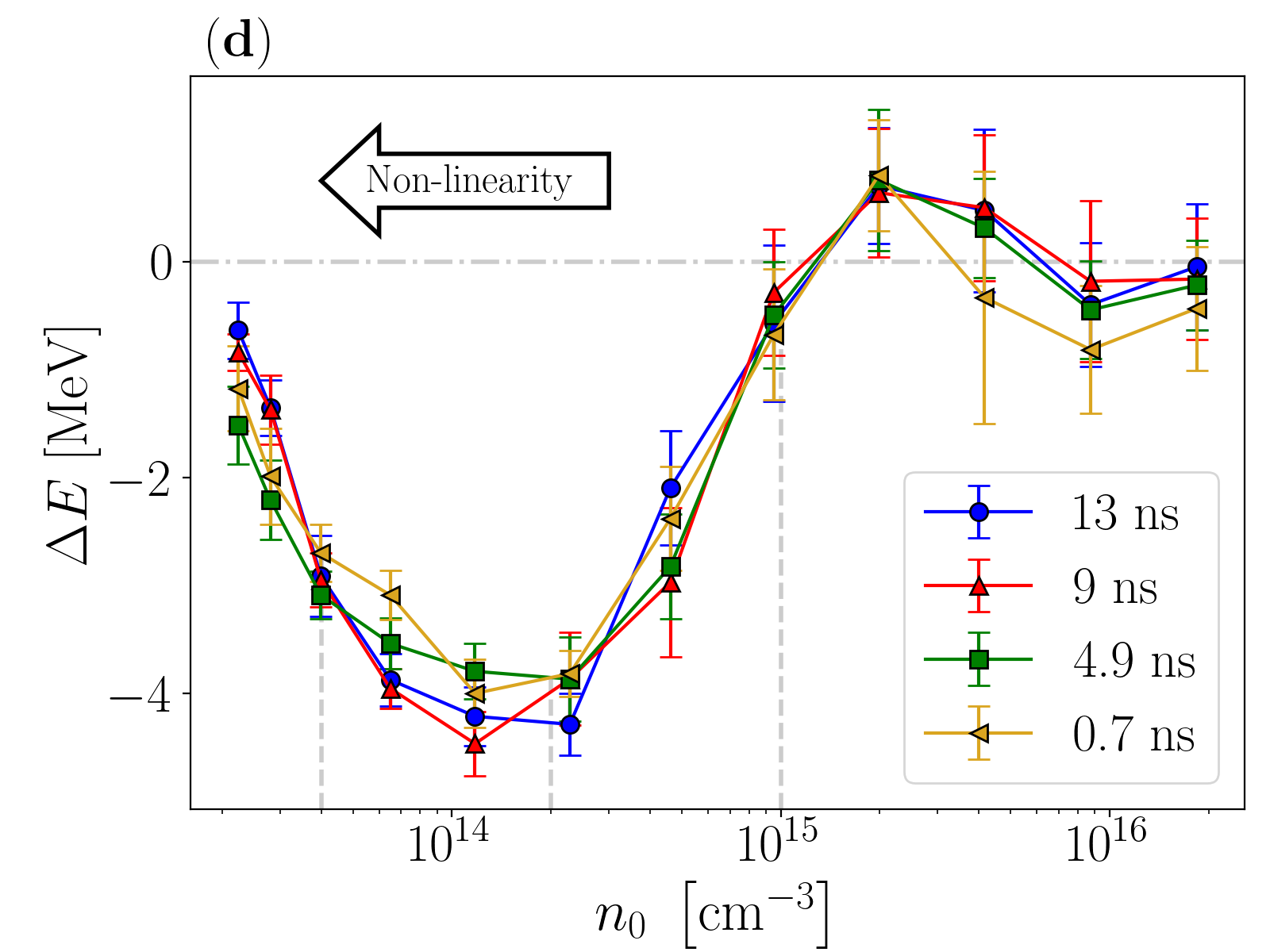}
    \caption{\small 
     (a) Sketch of the pump-and-probe configuration for the recovery time experiment performed in~\cite{pompili2024recoverytime}. A plasma wakefield is generated by a charged bunch (pump) in a plasma with density $n_0$. Another charged bunch (probe) enters the plasma after a time delay $\Delta t$. (b) The probe energy, $E_{\rm{P}}$, is measured downstream of the channel. (c) Additional measurement as in (a)-(b), but with a plasma unperturbed by the pump, resulting in a probe energy $E_{\rm{U}}$. 
     (d) experimental results on the characterization of $\Delta E=E_{\rm{P}}-E_{\rm{U}}$ as a function of $n_0$ for different values of the delay time $\Delta t$. Vertical dashed lines mark selected values of $n_0$ for which plasma waves are analyzed in Fig.~\ref{fig:splot_fluid_vs_pic}. The dashed-dotted horizontal line represents $\Delta E = 0$.
}\label{fig:sketch_ion_accumulation}
\end{figure*}

To explore the plasma recovery time issue, a proof-of-principle experiment has been conducted at SPARC\_LAB facility~\cite{ferrario2013sparclab} and a sub-nanosecond recovery time of a hydrogen plasma has been demonstrated via a a pump-and-probe configuration~\cite{pompili2024recoverytime} (see Fig.~\cref{fig:sketch_ion_accumulation} for a sketch): an electron bunch (pump) enters a plasma channel and generates a plasma wakefield perturbation, while a second electron bunch (probe) enters the plasma channel at a given time delay after the pump aimed at probing the perturbation induced by the pump. To explain the deceleration experienced by the probe, the authors invoked physical mechanisms rooted in the motion of ions triggered by the pump~\cite{khudiakov-2022, sahai2017ionmotion,zgadzaj-2020,DArcy2022,Gilljohann2019ionmotion} and used numerical simulations to support this claim. The numerical model used in~\cite{pompili2024recoverytime} consists of a simplified collisional model featuring the evolution of ions and electrons in the blowout regime~\cite{lu-2006}, without considering instabilities triggered by ion motion on the spatially resolved plasma waves~\cite{gorbunov2001plasma,gorbunov2003ionmotion, viera2012ionmotion,khudiakov-2022, sahai2017ionmotion,zgadzaj-2020}. While such a description may adapt to the late stages of plasma dynamics, the early stages need further scrutiny. This motivates the present complementary study, which aims to provide insight into the early stages of plasma dynamics by performing spatially resolved numerical simulations with collisionless plasma dynamics in the framework of the SPARC\_LAB experiment~\cite{pompili2024recoverytime}. \\ 

The paper is organized as follows: in Sec.~\ref{sec:experiment} an overview of the SPARC\_LAB experiment~\cite{pompili2024recoverytime} is given; in Sec.~\ref{sec:numerical_models} we summarize the relevant features of the numerical models used for simulations; numerical results are discussed in Sec.~\ref{sec:results}; conclusions are drawn in Sec.~\ref{sec:conclusions}. 

\section{Review on experiment}\label{sec:experiment}

The experiment~\cite{pompili2024recoverytime} uses a pump-and-probe configuration to study the plasma recovery time (see Fig.~\ref{fig:sketch_ion_accumulation}). An electron bunch (pump) enters a plasma with density $n_0$ and creates a wakefield; a second electron bunch (probe) enters the plasma at a time delay $\Delta t$ after the pump (Fig.~\ref{fig:sketch_ion_accumulation}(a)). The probe energy is then measured after the interaction with the perturbed plasma, $E_{\rm{P}}$, downstream of the capillary (Fig.~\ref{fig:sketch_ion_accumulation}(b)). An additional measurement was performed with a plasma unperturbed by the pump (Fig.~\ref{fig:sketch_ion_accumulation}(c)), resulting in a probe energy $E_{\rm{U}}$. The energy difference  $\Delta E=E_{\rm{P}}-E_{\rm{U}}$ is then studied to retrieve information on the plasma recovery time at changing $n_0$ and $\Delta t$: for a given plasma density $n_0$ and time delay $\Delta t$, $\Delta E \approx 0$ indicates that the plasma has recovered its initial properties; $\Delta E \neq 0$, instead, indicates that the probe has interacted with the plasma wake generated by the pump, hence after such $\Delta t$ the plasma has not recovered its initial state yet.

The experiment was performed using electron bunches (both pump and probe) with $500~\mbox{pC}$ charge, $84\pm0.1$~MeV energy, $167\pm 14$~fs duration (longitudinal spot size $\sigma_{\rm{z}} = 50\pm 4~\mu \rm{m}$), and transversal spot size $\sigma_{\rm{r}}=43\pm 2~\mu \rm{m}$ at the plasma entrance, resulting in a density that is $n_{\rm{b}} = 2\times 10^{15}~\rm{cm}^{-3}$. The bunch spot size was measured using a transition radiation screen installed at the capillary entrance, while the bunch duration and energy were measured with a radio frequency deflector and a magnetic spectrometer installed downstream of the capillary. The errors are computed as the standard deviations of 50 consecutive shots. For details on the experimental setup see~\cite{pompili2024recoverytime}.

Data from the experiment were collected for several pump-probe delays in the range $\Delta t=0.7-13$~ns. The energy difference $\Delta E$ as a function of the plasma density $n_0$ shows a non-monotonic trend with little dependency on the time delay $\Delta t$~(Fig.~\ref{fig:sketch_ion_accumulation}(d)). For plasma densities larger than the pump density (i.e. $n_0 \ge 2\times10^{15}~\mbox{cm}^{-3}$) the plasma response is closer to the linear regime and it is observed that $\Delta E \approx 0$; on the contrary, at lower plasma densities the plasma response is non-linear, and it is observed that the plasma probe experiences decelerating perturbations induced by the pump ($\Delta E < 0$) that are still present at $\Delta t = 13~\mbox{ns}$. The largest difference $\Delta E \approx -4$ MeV is observed with a plasma density $n_0 \approx 10^{14}~\rm{cm}^{-3}$. For values $n_0<10^{14}~\rm{cm}^{-3}$, the difference decreases again towards values closer to zero.

The hypothesis made in~\cite{pompili2024recoverytime} is that ions are pinched along the pump trajectories; they consequently attract electrons as well, so the probe experiences a higher plasma density, resulting in greater deceleration. However, no instruments or diagnostics to observe the evolution of ions inside the plasma capillary were available, like shadowgraphy techniques~\cite{shwab2013shadowgraphy, downer2018diagnostic}; hence, the authors in~\cite{pompili2024recoverytime} used numerical simulations to support the experimental observations. A simplified numerical method was used to reproduce electron and ion motion using a single equation for both, based on the non-linear blowout theory~\cite{lu-2006}, including collisional and thermal-pressure contributions. In the model, ions displacement is triggered by the electric field of the pump, and their dynamics are followed in time. At long times, an ion column is observed to persist when the pump density is larger than the background plasma density.

Experimental results reported Fig.~\ref{fig:sketch_ion_accumulation}(d) show that the non-monotonic behavior of $\Delta E$ as a function of the plasma density $n_0$ essentially persists for all $\Delta t$ studied, even the smallest. This suggests that some relevant information to support the experimental results may be retrieved from the analysis in the early stages of plasma dynamics. However, while the model used in~\cite{pompili2024recoverytime} is an appropriate tool to investigate late stages of ion motion triggered by the electric field of the pump after the electron plasma wake has dumped out, a rather different modeling needs to be adopted to model the early stages of plasma dynamics, where relevant physical time scales are very short and particle dynamics is essentially collisionless~\cite{macchi2013superintense, chen2016plasmaphysics}. Moreover, beyond the effect of the charged pump on the background ions~\cite{rosenzweig2005effects,gholizadeh-2011}, other contributions from the ponderomotive force of the oscillating fields of the plasma wave are expected~\cite{gorbunov2001plasma,gorbunov2003ionmotion, viera2012ionmotion, vieira2014ion}. The ion channel formation has a non-trivial impact on the plasma waves, leading to phase mixing and wave-breaking~\cite{gorbunov2001plasma,gorbunov2003ionmotion}, hence the plasma waves may destabilize before the time when particle collisions take place. All in all, it is unclear if and how the dynamics in the early stages influence the on-axis ion accumulation and if this can be reflected in the observations on the late stage dynamics summarized in Fig.~\ref{fig:sketch_ion_accumulation}(d). These considerations set a compelling case for the present complementary study on the early stages of plasma dynamics based on spatially resolved collisionless models, featuring plasma wave instabilities and embedding both the effect of the ion pinching set by the pump as well as the effects set by the ponderomotive forces of the electron plasma waves on ion motion.

\section{Numerical Simulation models with ion motion\label{sec:numerical_models}}
\begin{figure*}[t!]
    \centering
    \includegraphics[width=1.0\linewidth]{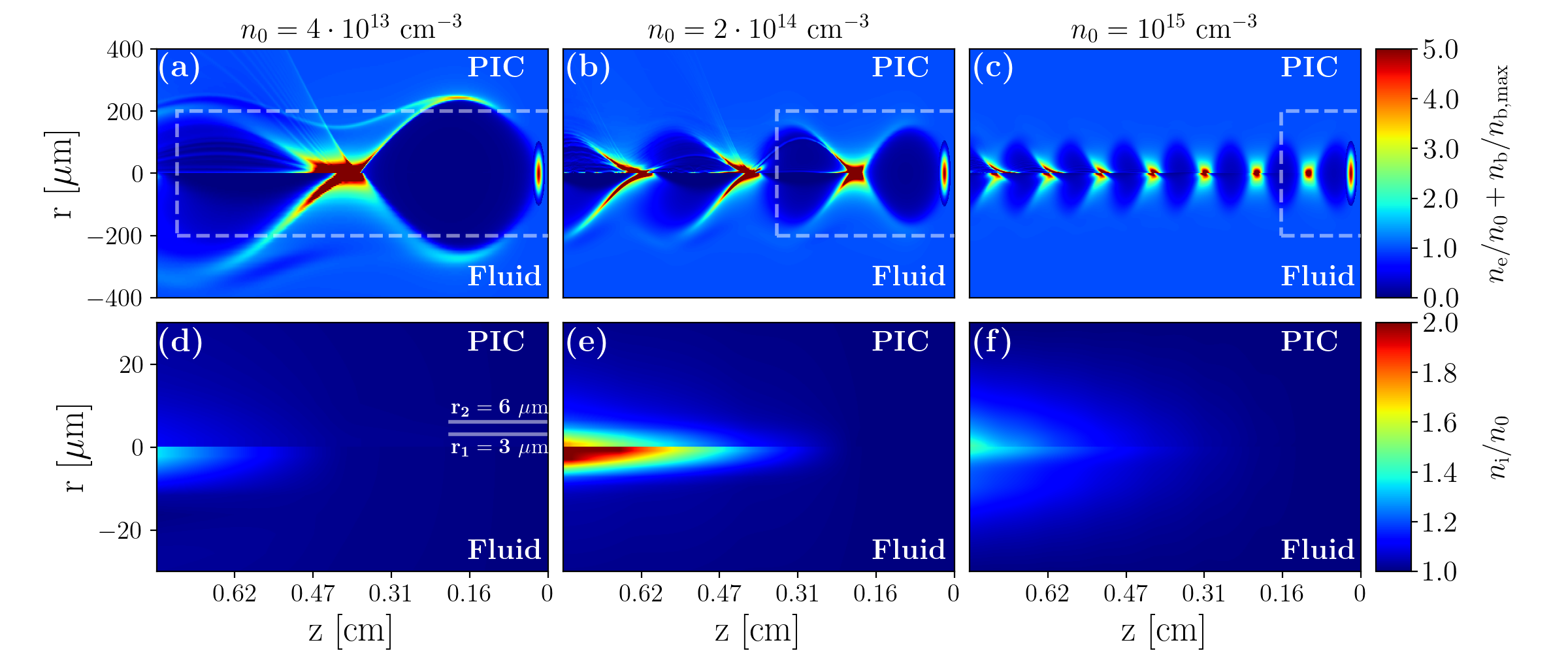}
    \caption{\small  color map in the $\rm{z}-\rm{r}$ plane for the normalized electron density $n_{\rm{e}}/n_0$ and bunch pump density $n_{\rm{b}}/n_{\rm{b,max}}$ ((a)-(c)), and for ion density $n_{\rm{i}}/n_0$ ((d)-(f)) after $0.78~\mbox{mm}$ ($~\sim 0.025~\rm{ns}$) of propagation of the pump into the plasma, moving from left to right. We report both data from fluid  (bottom-half) and PIC (top-half) simulations for three different values of the initial unperturbed plasma density $n_0$ (see dashed vertical lines in Fig.~\ref{fig:sketch_ion_accumulation}(d)). The dashed boxes represent the window where the maximum of the focusing field, $W_{\rm{r,max}}=E_{\rm{r}} - cB_{\rm{\phi}}$, analyzed in Fig.~\ref{fig:wave-breaking_estimation} is computed. Horizontal segments in (d) represent the values of $r$ used to compute ion density analyzed in Fig.~\ref{fig:ions_density_non_monodromy_plot}.}
\label{fig:splot_fluid_vs_pic}
\end{figure*}
In the plasma acceleration community, spatially resolved plasmas are commonly simulated using numerical codes based on two main classes of physical models: Particle-in-Cell (PIC) models and fluid models. Both are rooted in kinetic theory, since the reference equations governing plasma dynamics are given by the Vlasov–Maxwell system~\cite{birdsall1991pic, cercignani2002relativistic}:
\begin{align} \label{eq:vlasov}
    &\frac{d f_{\rm{s}}}{d t} + q_{\rm{s}} \left( \mathbf{E} + \mathbf{v}_{\rm{s}}\times\mathbf{B} \right)\cdot \frac{\partial f_{\rm{s}}}{\partial \mathbf{p}_{\rm{s}}} = 0~,
\end{align}
where the subscript $\rm{s}$ indicates the plasma species (i.e., $\rm{s=i}$ for ions and $\rm{s=e}$ for electrons), $f_{\rm{s}}$ is the distribution function representing the probability density to find a particle in the volume element $d\mathbf{x}_{\rm{s}}d\mathbf{v}_{\rm{s}}$, $q_{\rm{s}}$ is the species charge ($q_{\rm{e}}=-e$ for the electrons and $q_{\rm{i}} = e$ for the ions, assuming a hydrogen plasma, where $e$ is the magnitude of the electron charge), $\mathbf{v}_{\rm{s}}$ is the microscopic velocity of the species, $\mathbf{p}_{\rm{s}}=m_s\gamma(\mathbf{v}_{\rm{s}})\mathbf{v}_{\rm{s}}$ the microscopic momentum and $\gamma(\mathbf{v}_{\rm{s}})=1/\sqrt{1-(\mybf{v}_{\rm{s}}/c)^2}$ is the associated Lorentz factor, $c$ is the speed of light, $m_{\rm{s}}$ is the species mass, $\mathbf{E}$ and $\mathbf{B}$ are the electric and the magnetic field, respectively, which evolve following Maxwell's equations: 
\begin{equation}\label{eq:maxwell} 
    \begin{aligned}
    &\nabla \cdot \mybf{E} = \frac{\sum_{\rm{s}} q_{\rm{s}} n_{\rm{s}}}{\epsilon_0}~,\\
    &\nabla \cdot \mybf{B} = 0~,\\
    &\nabla \times \mybf{E} = -\partialt{\mybf{B}}\\
    & \nabla \times \mybf{B} =\mu_0 (- e n_{\rm{b}} \mybf{u}_{\rm{b}} + \sum_{\rm{s}} q_{\rm{s}} n_{\rm{s}} \mybf{u}_{\rm{s}})+ \epsilon_0 \mu_0 \partialt{\mybf{E}} \;,
    \end{aligned}
\end{equation}
where $\mu_0, \epsilon_0$ are the magnetic permeability and electric permittivity in vacuum, $(n_{\rm{s}}, n_{\rm{b}})$ and $(\mybf{u}_{\rm{s}}, \mybf{u}_{\rm{b}})$ are the number density and macroscopic velocity of plasma species and pump, respectively.
PIC models~\cite{lehe2016fbpic,vay2018WarpX,davidson2015osiris,Mehrling2014hipace,benedetti2008aladyn}, which describe plasma evolution by resolving particle dynamics at the microscopic level within a continuum kinetic phase space, represent the state of the art in the field~\cite{birdsall1991pic}, as they are able to capture most of the relevant plasma physics. However, due to their microscopic nature, achieving this level of accuracy often entails a high computational cost, particularly to minimize numerical noise. On the other hand, fluid models~\cite{parise-2022,Tomassini2016qfluid,benedetti2010inferno} describe plasma evolution through macroscopic equations for density and momentum, obtained by coarse-graining the kinetic equations of the Vlasov–Maxwell system (Eqs.~\eqref{eq:vlasov}-\eqref{eq:maxwell}) and closing the corresponding moment hierarchy with a suitable closure scheme. While fluid models cannot capture all kinetic effects~\cite{boyd2003kineticignoredfluid}, they remain valid for many experimental conditions of interest, providing fast and accurate results that are free of numerical noise due to particle undersampling.

In both approaches, a cold plasma approximation is used, since the temperatures of the plasma at rest are in the $1 - 10~\rm{eV}$ range~\cite{anania2014initialplasmatemperature,gonsalves2019initialplasmatemperature}, much lower than the electron rest energy ($\sim 0.511$ MeV). Since PIC models represent the state of the art of numerical simulations in the plasma acceleration community~\cite{vay2016picreviw,lindstrom2025pwfareview, esarey2009lasersrev}, they are taken as a ground truth for assessing the range of validity of fluid-model predictions.

The comparison between PIC and fluid models in the cold case~\cite{massimo2016comparisons, parise-2022} and in the warm case~\cite{simeoni2024thermalLB,simeoni2024thermalfluid, simeoni2025thermalwakestructure} has been addressed in the literature. In all these works, however, ion motion was neglected. In this present study, we take a step forward in this direction by performing a systematic comparison between the two approaches in the presence of self-consistent ion dynamics.
\subsection{Particle-in-Cell (PIC) models}
PIC models~\cite{vay2018WarpX,davidson2015osiris,Mehrling2014hipace,benedetti2008aladyn} provide a bottom-up solution to the Vlasov-Maxwell system by evolving individual Lagrangian particles according to the Newton-Lorentz equations~\cite{birdsall1991pic,vay2016picreviw}:
\begin{equation}
    \begin{split}\label{eq:micro_pic}
    &\frac{d\mybf{x}_{\rm{s}}}{dt} = \mybf{v}_{\rm{s}} ~,  \\
    &\frac{d\mybf{p}_{\rm{s}}}{dt} = q_{\rm{s}} \left( \mybf{E}+\mybf{v}_{\rm{s}} \times \mybf{B} \right) ~, \\
    &\mybf{p}_{\rm{s}}   = m_{\rm{s}} \gamma\left(\mybf{v}_{\rm{s}}\right) \mybf{v}_{\rm{s}}~.\\ 
    \end{split}
\end{equation}
Solving this set of equations coupled with Eqs.~\cref{eq:maxwell} for each individual plasma particle is computationally prohibitive even on modern high-performance computing facilities: for this reason, PICs employ macro-particles, each representing an ensemble of particles with the same position and velocity. This approximation introduces numerical noise, which is the main source of uncertainty. In this work, we used the code FBPIC~\cite{lehe2016fbpic}, a quasi-3D PIC code that solves Eqs.~\cref{eq:micro_pic} in 3D with a leapfrog scheme following~\cite{vay2008pusher} while Eqs.~\cref{eq:maxwell} are solved with a Pseudo Spectral Analytical Time Domain (PSATD) algorithm~\cite{buneman19803demparticles} taking advantage of the cylindrical axial symmetry to spare computational time and using azimuthal Fourier decomposition to recover 3D features for the electromagnetic fields. For our simulations, we used a complete cylindrical axially symmetric framework with only the zero azimuthal mode. 

\subsection{Fluid models}
Fluid codes~\cite{vay2018WarpX,massimo2016comparisons, Tomassini2016qfluid} solve macroscopic equations for the plasma evolution:
\begin{equation}
    \label{eq:rel-euler}
    \begin{split}
    &\partialt{n_{\rm{s}}} + \nabla \cdot (n_{\rm{s}} \mathbf{u}_{\rm{s}}) = 0 \; , \\
    &\left( \partialt{} + \mybf{u}_{\rm{s}} \cdot \nabla \right)\left(m_{\rm{s}} \gamma\left(\mybf{u}_{\rm{s}}\right) \mybf{u}_{\rm{s}}\right) 
    = q_{\rm{s}} (\mybf{E} + \mybf{u}_{\rm{s}} \times \mybf{B}) \; ,
    \end{split}
\end{equation}
coupled with the Maxwell Eqs.~\cref{eq:maxwell}. The set of Eqs.\cref{eq:rel-euler} offers the advantage of being free from numerical noise coming from particle undersampling, at the expense of neglecting certain kinetic effects~\cite{boyd2003kineticignoredfluid}. This characteristic arises from the fluid formulation itself, which is derived from a hierarchy of conservation equations obtained by integrating Eq.~\cref{eq:vlasov} over the kinetic velocity space~\cite{cercignani2002relativistic, rezzolla-2013}. In this work, we used a Finite Difference Time Domain (FDTD) scheme with Yee cell~\cite{yee-1966} and the Lattice Boltzmann Method (LBM) described in detail in~\cite{parise-2022,simeoni2024thermalLB, simeoni2025thermalwakestructure} to solve Eqs.~\cref{eq:rel-euler,eq:maxwell} in cylindrical axial symmetry. LBM includes an additional tunable diffusion term that we keep as small as the method's stability permits. 
\section{Numerical Results \label{sec:results}}
\begin{figure}[!t]
    \centering
\includegraphics[width=0.7\textwidth]{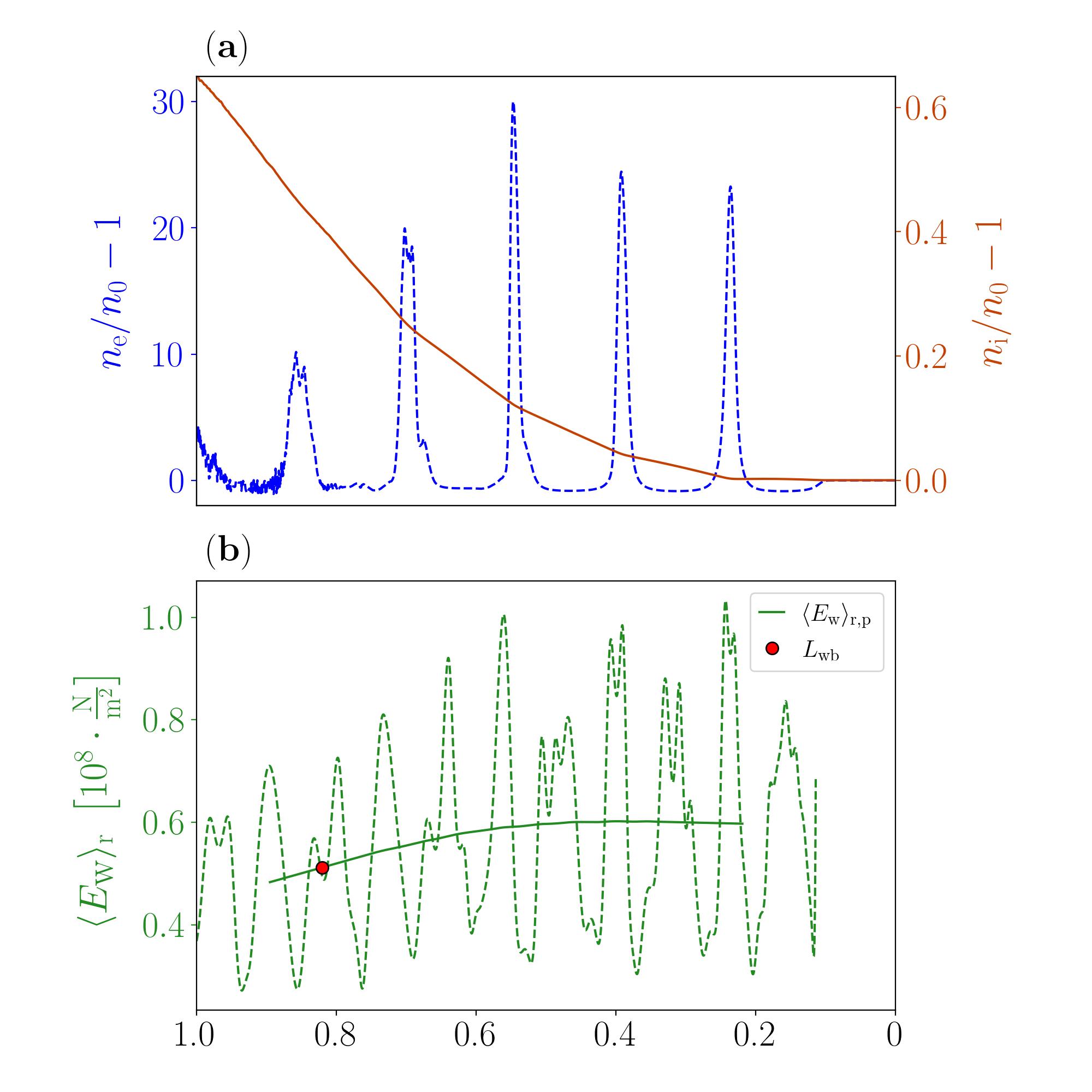} 
    \caption{\small Data from simulations with an initial plasma density $n_0 = 4 \cdot 10^{14}~\rm{cm}^{-3}$. (a) Normalized electron (blue line) and ion (yellow line) density as a function of $z$, computed at $r=3~\rm{\mu m}$ from the axis. (b) Radial integration of the electromagnetic energy, $\langle E_{\rm{W}} \rangle_{\rm{r}}$ (dashed green line), and its mean value over a plasma period, $\langle E_{\rm{W}}\rangle_{\rm{r,p}}$ (solid green line), as a function of $z$. The wave-breaking point is indicated with a red bullet.} \label{fig:wave-breaking_estimation_sketch}
\end{figure}
We performed simulations in the early stages of plasma dynamics ($< 0.1~\rm{ns}$) studying the plasma response for different background plasma densities $n_0$ ranging from $2\cdot10^{13}~\mbox{cm}^{-3}$ to $1 \cdot 10^{16}~\mbox{cm}^{-3}$. For each value of $n_0$, we started from a plasma at rest further perturbed by a rigid pump moving at the speed of light $c$ (direction from left to right in ~Figs.~\ref{fig:splot_fluid_vs_pic},~\ref{fig:wave-breaking_estimation_sketch}). The pump is modelled with an axially symmetric Gaussian density distribution with the same charge and rms widths $\sigma_{\rm{r},\rm{z}}$ as those of the experiment (see Sec.~\ref{sec:experiment}). The simulation box in cylindrical axially symmetric coordinates is a two dimensional domain $L_{\rm{z}} \times L_{\rm{r}}$ with longitudinal length $L_{\rm z} = 0.78~\mbox{cm}$ ($\sim 0.025~\mbox{ns}$) and transversal length $L_{\rm r} = 1~\mbox{mm}$ discretized with a uniform resolution $\Delta z = \Delta r = 0.005 / k_{\rm{p}} $, dependent on the cold plasma wave-number $k_{\rm{p}} = \omega_{\rm{p}}/c$, where $\omega_{\rm{p}} = \sqrt{n_0 e^2 / (m_{\rm{e}} \epsilon_0)}$ is the cold plasma frequency. The time step was set to $\Delta t = 0.0025/\omega_{\rm{p}}$ for the fluid code and to $\Delta t = 0.005/\omega_{\rm{p}}$ for the PIC code. Following a resolution study, the PIC simulations were performed using a number of particles per cell ranging from 25 to 64, as no significant differences were observed in the results when varying this parameter within this range. Simulations implement a moving window technique~\cite{birdsall1991pic}. The pump is initialized in vacuum, and a short linear plasma ramp is implemented to avoid numerical injection events. The ion mass is set to proton mass, assuming a completely ionized hydrogen plasma. Boundary conditions are open on the left, right, and upper boundaries, while axial symmetry is imposed on the bottom boundary. For details on the implementation of boundary conditions see~\cite{parise-2022, simeoni2024thermalLB} and~\cite{lehe2016fbpic}, for the fluid models and PIC, respectively.

\begin{figure}[!t]
    \centering
\includegraphics[width=0.7\textwidth]{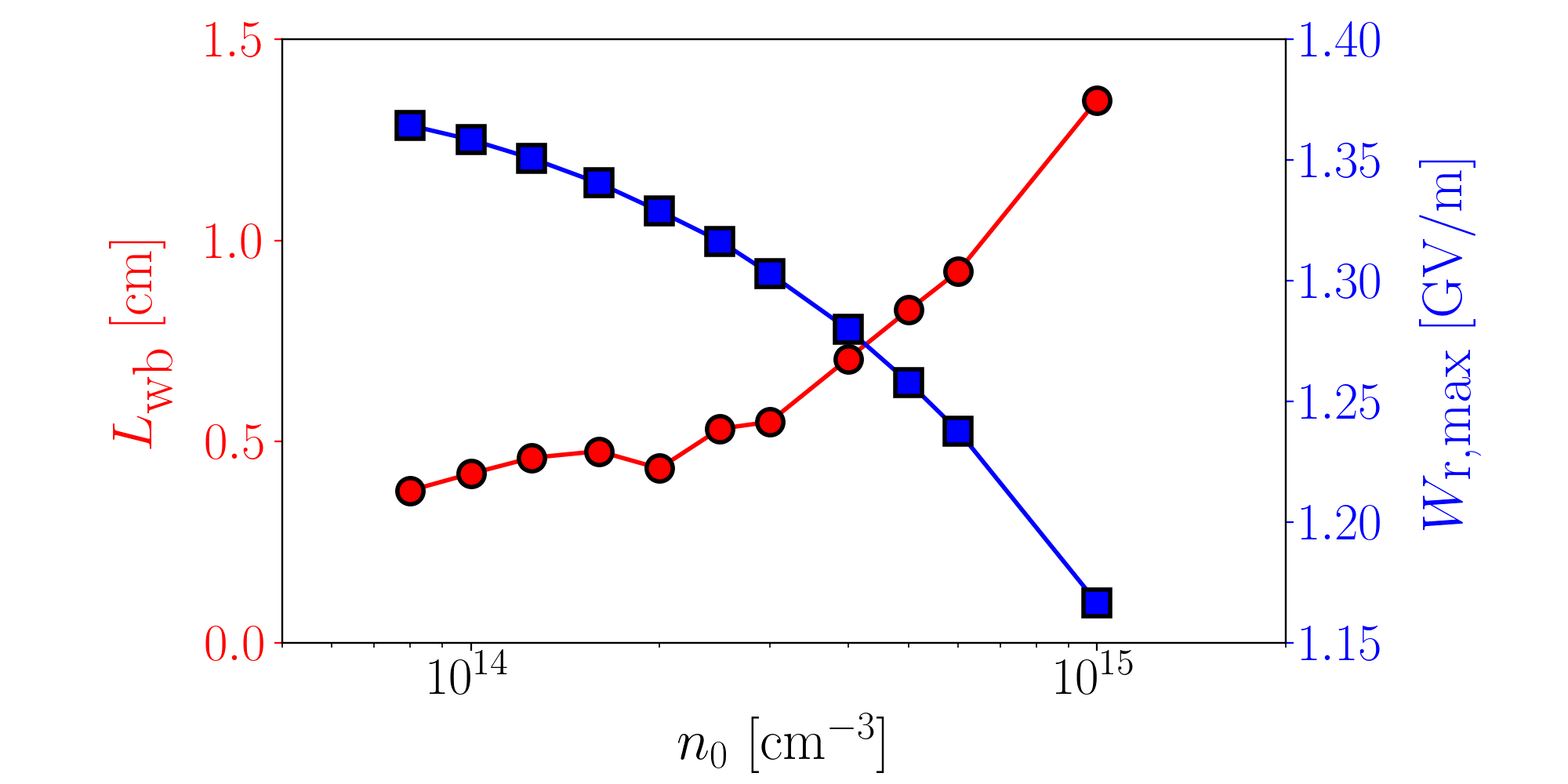}
    \caption{ \small Electron wave-breaking length, $L_{\rm{wb}}$ (red dots), and maximum of the focusing wakefield, $W_{\rm{r,max}}=E_{\rm{r}} - cB_{\rm{\phi}}$ (blue squares), as a function of the initial plasma density $n_0$, computed from PIC simulations (see text for details).}   \label{fig:wave-breaking_estimation}
\end{figure}

In Fig.~\ref{fig:splot_fluid_vs_pic}, we show color maps of electron and ion density, both from PIC (top half) and fluid (bottom half) simulations. We choose three values of initial plasma density $n_0=4 \cdot 10^{13}, \,2  \cdot 10^{14},\, 10^{15}~\mbox{cm}^{-3}$ (see dashed vertical lines in Fig.~\ref{fig:sketch_ion_accumulation}) to highlight the different regimes that the plasma wave develops under the pump perturbation. 
Results for the electron plasma wave for both fluid and PIC models (Fig.~\ref{fig:splot_fluid_vs_pic}(a)-(c)) show typical common features: the plasma wave is closer to linear regime for large $n_0$ and becomes progressively non-linear as $n_0$ decreases, hence as $n_0$ decreases the plasma oscillation becomes less frequent and the electron bubble grows in size~\cite{lu2010bubblesizescaling,lindstrom2025pwfareview}. Electrons expelled away from the axis due to the interaction with the pump favour the emergence of the ion cavity (bubble); expelled electrons are then attracted back, forming a sheath around the bubble and a region of high density in the tail of the bubble~\cite{lu-2006, simeoni2025thermalwakestructure, esarey2009lasersrev, lindstrom2025pwfareview}. As expected~\cite{massimo2016comparisons,simeoni2025thermalwakestructure} the bubble sizes observed for fluid and PIC models are well in agreement; the sheath around the bubble is more prominent for non-linear regimes (see Fig.~\ref{fig:splot_fluid_vs_pic}(a)), is present in both fluid and PIC descriptions although smoother for the fluid case. Regarding ion motion  (Fig.~\ref{fig:splot_fluid_vs_pic}(d)-(f)), a cone-like structure in the ion distribution appears in the tail of the wake with a radial spatial extent that does not sensibly depend on the model used nor on $n_0$, although $n_0$ is changed by orders of magnitude, indicating that its main dependence is in the pump structure that is kept unchanged in our simulations: this well agrees with previous observations on ion channel formation triggered by laser pulses~\cite{gorbunov2001plasma,gorbunov2003ionmotion}. We observe differences in the magnitude of the on-axis ion density from fluid and PIC models, with the density predicted by the fluid models being larger (Fig.~\ref{fig:splot_fluid_vs_pic}(e)). An important feature emerging from the analysis of the electron plasma waves in Fig.~\ref{fig:splot_fluid_vs_pic}(a)-(c), is that plasma oscillations progressively loose coherence: this is particularly evident in the analysis of the structure of the bubble tail, wherein 
the electron density develops a curved backward bending away from the axis~\cite{vieira2014ion}  as the distance from the pump increases (Fig.~\ref{fig:splot_fluid_vs_pic}(a)-(c)). These are the typical signatures of fine-scale mixing and wave-breaking, whose emergence is known to be accompanied by a disagreement between fluid and PIC models, especially close to the axis in the tail of the bubble where electron density is larger~\cite{gorbunov2001plasma,gorbunov2003ionmotion,vieira2014ion}. Indeed, fluid descriptions cannot capture all the kinetic effects involved in the electrons evolutions, a fact that is particularly evident in the analysis of the high density regions in the bubble tails in non-linear regimes (Fig.~\ref{fig:splot_fluid_vs_pic}(a)-(b)), where we observe fast electrons escaping from the wave orbit in the PIC model but not in the fluid counterpart. To make the discussion on wave-breaking more quantitative, we analyzed the results from the PIC simulations based on the considerations made in~\cite{gorbunov2001plasma,gorbunov2003ionmotion}. The emergence of wave-breaking implies that a part of the electromagnetic energy density of the wave, $E_{\rm{W}} = \epsilon_0 |\mybf{E}|^2/2 + |\mybf{B}|^2/(2\mu_0)$, is transferred to fast electrons and ions. Hence, we look at the integral over the transversal cylindrical coordinate of the electromagnetic energy density, $\langle E_{\rm W} \rangle_{\rm{r}}(z)=\int_{0}^{L_{\rm r}} rE_{\rm{W}}(z,r) \,dr$, and further average over a plasma period, $\langle E_{\rm W} \rangle_{\rm{r},\rm{p}}(z)=\int_{z-\pi/k_{\rm{p}}}^{z+\pi/k_{\rm{p}}}  \langle E_{\rm W} \rangle_{\rm{r}}(z')\,dz'$. The wake-breaking length scale $L_{\rm{wb}}$ is defined as the distance from the pump at which $\langle E_{\rm{W}} \rangle_{\rm{r},\rm{p}}$ is decreased by 15\% (see Fig.~\ref{fig:wave-breaking_estimation_sketch}). We have verified that a change from 15\% to larger fractions does not change the qualitative nature of our results. In Fig.~\ref{fig:wave-breaking_estimation} we show $L_{\rm{wb}}$ as a function of $n_0$. Notice that we extended the simulation box to $1.3~\rm{cm}$ for the simulation at $n_0=10^{15}~\rm{cm}^{-3}$ to reach the wave-breaking point for our estimation algorithm. For completeness, we also analyzed the importance of the ponderomotive force of the electron wave, which is another important actor in promoting on-axis ion accumulation~\cite{gorbunov2001plasma,gorbunov2003ionmotion}. Although the ponderomotive force is intrinsically related to the spatial gradient of the wave electromagnetic energy, we chose to characterize its influence indirectly by reporting the maximum of the focusing wakefield, motivated by the fact that the peak focusing field provides a reliable alternative for the effective strength of the ponderomotive force acting on the ions, while avoiding additional complexity associated with direct gradient evaluations. In Fig.~\ref{fig:wave-breaking_estimation}, we report the maximum of the focusing wakefield that we found in a box of transversal size of $200~\rm{\mu m}$ (dashed box in Fig.~\ref{fig:splot_fluid_vs_pic}).

\begin{figure}[t!]
    \centering
    \includegraphics[width=0.7\linewidth]{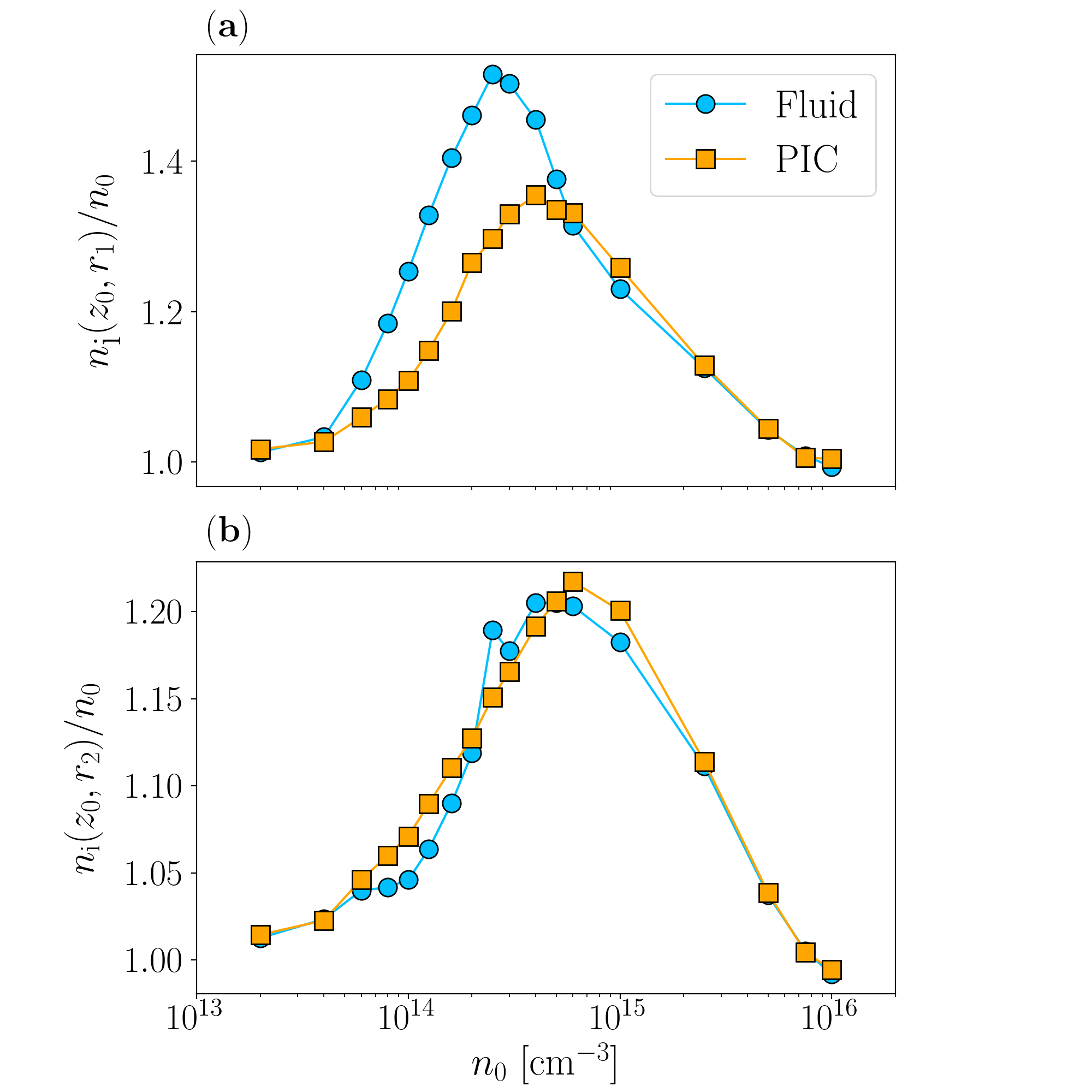}
    \caption{\small The normalized ion density $n_{\rm{i}}/n_0$ as a function of $n_0$. We report data for both the PIC (yellow squares) and fluid (light blue circles) simulations, at a fixed value of the longitudinal coordinate $z_0 = 0.78~\rm{cm}$ and for two different values of transversal coordinate: $r_1 = 3~\mu \rm{m}$ (a) and $r_2=6~\mu \rm{m}$ (b).}  \label{fig:ions_density_non_monodromy_plot}
\end{figure}
Both the ponderomotive force~\cite{gorbunov2001plasma,gorbunov2003ionmotion, viera2012ionmotion, vieira2014ion} and the attraction of the pump~\cite{rosenzweig2005effects,gholizadeh-2011,pompili2024recoverytime} promote on-axis ion accumulation and both effects increase as $n_0$ decreases. As $n_0$ decreases, however, the wave-breaking lengthscale $L_{\text wb}$ gets smaller (see Fig.~\ref{fig:wave-breaking_estimation}) and the plasma wave breaks earlier, hence the ponderomotive force can last for a shorter period. Thus, as $n_0$ decreases, the force promoting on-axis ion accumulation experiences a delicate balance between increasing intensity and decreasing duration, which one expects to give rise to a non-monotonic behavior in the magnitude of the on-axis ion density. These conclusions are supported by the analysis shown in  Fig.~\ref{fig:splot_fluid_vs_pic}(d)-(f), where we indeed observe that in both fluid and PIC models, the magnitude of the on-axis ion accumulation is compatible with a non-monotonic behavior, being maximum for the snapshot reported in Fig.~\ref{fig:splot_fluid_vs_pic}(e). This well echoes the experimental observations on $\Delta E$ (see Fig.~\ref{fig:sketch_ion_accumulation}(d)). To address this issue more quantitatively, in Fig.~\ref{fig:ions_density_non_monodromy_plot} we report the normalized ion density, $n_{\rm{i}}/n_0$, as a function of $n_0$. The ion density is taken from two single points at the edge of the simulation boundary, $z_0=0.78~\rm{cm}$, and for two different radial coordinates $r_1=3~\mu\rm{m}$ and $r_2=6~\mu \rm{m}$ (see horizontal segments in Fig.~\ref{fig:splot_fluid_vs_pic}(d)). Closer to the axis ($r_1=3~\mu \rm{m}$) results of fluid models overestimate PIC results, especially in non-linear regimes ($n_0 \le 10^{15}~\mbox{cm}^{-3}$); more away from the axis ($r_2=6~\mu \rm{m}$) the agreement between the two models considerably improves. Overall, a maximum in ion density is observed when $n_0 \approx 4-5 \cdot 10^{14}~\mbox{cm}^{-3}$, overestimating the value of $n_0 \approx 1-2 \cdot 10^{14}~\mbox{cm}^{-3}$ where experimental results find the largest difference $\Delta E$. This discrepancy could be attributed to different factors. While the numerical results pertain the early stages of plasma dynamics until about $0.025~\rm{ns}$, the first dataset in the experiment is taken for $0.7~\rm{ns}$, thus there is additional dynamics after wave-breaking that we have not simulated in details and could have an impact on the value of $n_0$ for which the probe deceleration induced by the perturbed plasma is maximized. While fluid models lose their validity after wave-breaking, PIC simulations could still be used, but become noisy (see Fig.~\ref{fig:wave-breaking_estimation_sketch}(a)). Moreover, our numerical simulations have been performed with a rigid pump (i.e., infinite energy), while in experiments the pump has a finite energy, resulting in an initial focusing in the plasma ramp at the entrance of the capillary and a shape change during the interaction with the plasma. Finally, we note that in the presence of wave-breaking, the wave energy gets deposited in the plasma channel, leading to significant electron heating~\cite{khudiakov-2022}, implying that the adoption of warm models for simulation~\cite{simeoni2024thermalLB,simeoni2025thermalwakestructure,diederichs2023temperature} might have an impact.
\section{Conclusions \label{sec:conclusions}}
We performed numerical simulations to characterize plasma dynamics within the experimental framework of the recovery time experiment conducted at the SPARC\_LAB facility~\cite{pompili2024recoverytime}. The experiment explored the 
recovery time of a hydrogen plasma at the sub-nanosecond timescale with a pump-and-probe setup: a plasma with background density $n_0$ is perturbed by a charged pump, and the plasma perturbation is then tested with a charged probe sent into the plasma after the pump. The experiment reported a non-monotonic dependence of the probe deceleration as a function of $n_0$. On-axis ion accumulation triggered by the plasma waves has been invoked as the main cause for the pump deceleration. To complement these observations, we thus investigated the problem of ion dynamics in the same set-up of the SPARC\_LAB experiment in~\cite{pompili2024recoverytime} via numerical simulations based on both fluid~\cite{parise-2022} and PIC~\cite{lehe2016fbpic} models. On the one hand, our work helps elucidate the physical mechanisms underlying the non-monotonic behavior of the probe deceleration observed in experiments, showing that it arises from a delicate balance between the strength and duration of the forces triggering ion displacement. On the other hand, this allows for a quantitative assessment of fluid vs. PIC models in a context featuring ion dynamics, thus extending earlier studies ~\cite{massimo2016comparisons} that considered immobile ions.

Although the simulations reproduce a non-monotonic dependence of the on-axis ion density on $n_0$, thereby capturing the main qualitative trends observed experimentally, the remaining discrepancies indicate that the present modeling can be naturally improved by incorporating additional physical effects in the simulations. In particular, the evolution of the driving pump, which is treated here as rigid, is expected to influence the wakefield structure in realistic experimental conditions~\cite{lindstrom2025pwfareview}. In addition, the inclusion of finite plasma temperature represents another important ingredient, as recently highlighted by several studies~\cite{simeoni2025thermalwakestructure,simeoni2024thermalLB,simeoni2024thermalfluid}. Another promising direction for future work is the extension of the present analysis to longer-term plasma dynamics, thus promoting a direct comparison with the numerical results presented in~\cite{pompili2024recoverytime}. Such an extension raises the issue of selecting the most appropriate simulation strategy for treating these time scales. 
In Refs.~\cite{khudiakov-2022,zgadzaj-2020}, for example, the early-stage plasma evolution ($t<100~\rm{ps}$) is modeled using the PIC code OSIRIS~\cite{davidson2015osiris}, whose results are then employed as initial conditions for late-stage simulations ($t>100~\rm{ps}$) performed with the quasi-static code LCODE~\cite{sosedkin2016lcode}. 
As discussed therein, the late-stage evolution involves additional physical processes, such as collisions and thermodynamic effects, which are often neglected in standard PIC frameworks. 
In this context, fluid models may represent a viable and complementary alternative to quasi-static approaches~\cite{sosedkin2016lcode, Mehrling2014hipace, huang2006quickpic}, as they allow for the inclusion of such effects within a self-consistent macroscopic description~\cite{parise-2022, simeoni2024thermalLB}. A quantitative understanding of plasma recovery time is therefore essential for the development of high-repetition-rate plasma wakefield accelerators, as required by existing facilities and future light sources such as EuPRAXIA~\cite{pap21}.

\section*{Acknowledgments}
This work has received funding from the European Union’s Horizon Europe Research and Innovation program under Grant Agreement No. 101079773 (EuPRAXIA Preparatory Phase) and No. 101188004 (PACRI). This work was supported by the Italian Ministry of University and Research (MUR) under the FARE program (No. R2045J8XAW), project "Smart-HEART". MS gratefully acknowledges the support of the National Center for HPC, Big Data and Quantum Computing, Project CN\_00000013 - CUP E83C22003230001, Mission 4 Component 2 Investment 1.4, funded by the European Union - NextGenerationEU. Financial support from the project DYNAFLO (CUP E85F21004290005) of Tor Vergata University of Rome is acknowledged. MS, FG, DS and GP gratefully acknowledge Fabio Bonaccorso for his technical support. The authors acknowledge the SPARC\_LAB collaboration for useful discussions and suggestions.

\section*{ Statements and Declarations}

\subsection*{Conflict of Interests}
The authors declare that they have no conflict of interests.




\section*{Data Availability Statement}

The data that support the findings of this study are available from the corresponding author upon reasonable request.

\printbibliography

@article{Ackermann2007fel,
  title = { Operation of a free-electron laser from the extreme ultraviolet to the water window},
 author = {W. Ackermann and others},
  journal = {Nature Photonics},
  volume = {1},
  issue = {6},
  pages = {336–342},
  numpages = {6},
  year = {2007},
  month = {June},
  publisher = {Nature},
  doi = {10.1038/nphoton.2007.76},
  url = {https://doi.org/10.1038/nphoton.2007.76}
}

@article{anania2014initialplasmatemperature,
title = {Design of a plasma discharge circuit for particle wakefield acceleration},
journal = {Nuclear Instruments and Methods in Physics Research Section A: Accelerators, Spectrometers, Detectors and Associated Equipment},
volume = {740},
pages = {193-196},
year = {2014},
note = {$1^{st}$ European Advanced Accelerator Concepts Workshop 2013},
issn = {0168-9002},
doi = {https://doi.org/10.1016/j.nima.2013.10.053},
url = {https://www.sciencedirect.com/science/article/pii/S0168900213014344},
author = {M.P. Anania and others},
}

@article{Argyropoulos2018xband,
  title = {Design, fabrication, and high-gradient testing of an $X$-band, traveling-wave accelerating structure milled from copper halves},
  author = {Theodoros Argyropoulos and others},
  journal = {Physical Review Accelerators and Beams},
  volume = {21},
  issue = {6},
  pages = {061001},
  numpages = {11},
  year = {2018},
  month = {June},
  publisher = {American Physical Society},
  doi = {10.1103/PhysRevAccelBeams.21.061001},
  url = {https://link.aps.org/doi/10.1103/PhysRevAccelBeams.21.061001}
}

@article{pap21,
  title={EuPRAXIA Conceptual Design Report},
  author={Assmann, RW and others},
  journal={The European Physical Journal Special Topics},
  volume={229},
  number={24},
  pages={3675--4284},
  year={2020},
  publisher={Springer},
  doi={10.1140/epjst/e2020-000127-8}
}

@article{benedetti2010inferno,
    author = {Benedetti, C. and Schroeder, C. B. and Esarey, E. and Geddes, C. G. R. and Leemans, W. P.},
    title = {Efficient Modeling of Laser‐Plasma Accelerators with INF{\&}RNO},
    journal = {AIP Conference Proceedings},
    volume = {1299},
    number = {1},
    pages = {250-255},
    year = {2010},
    month = {November},
    issn = {0094-243X},
    doi = {10.1063/1.3520323},
    url = {10.1063/1.3520323},
}

@ARTICLE{benedetti2008aladyn,
  author={Benedetti, Carlo and Sgattoni, Andrea and Turchetti, Giorgio and Londrillo, Pasquale},
  journal={IEEE Transactions on Plasma Science}, 
  title={${\tt ALaDyn}$: A High-Accuracy PIC Code for the Maxwell–Vlasov Equations}, 
  year={2008},
  volume={36},
  number={4},
  pages={1790-1798},
  doi={10.1109/TPS.2008.927143}
}

@book{birdsall1991pic,
  title={Plasma Physics via Computer Simulation},
  author={C.K. Birdsall and A.B Langdon},
  year={1991},
  publisher={CRC press},
  DOI = {10.1201/9781315275048}
}

@article{buneman19803demparticles,
title = {Principles and capabilities of 3-D, E-M particle simulations},
journal = {Journal of Computational Physics},
volume = {38},
number = {1},
pages = {1-44},
year = {1980},
issn = {0021-9991},
doi = {https://doi.org/10.1016/0021-9991(80)90010-8},
url = {https://www.sciencedirect.com/science/article/pii/0021999180900108},
author = {O. Buneman and C.W. Barnes and J.C. Green and D.E. Nielsen}
}

@book{boyd2003kineticignoredfluid, 
title     = {The Physics of Plasmas}, 
publisher = {Cambridge University Press}, 
author    = {Boyd, T. J. M. and Sanderson, J. J.}, 
year      = {2003},
doi={doi.org/10.1017/CBO9780511755750}
}

@book{cercignani2002relativistic,
  title={The Relativistic Boltzmann Equation: Theory and Applications},
  author={Cercignani, C. and Kremer, G. M.},
  year={2002},
  publisher={Birkhäuser Basel},
  doi={10.1007/978-3-0348-8165-4}
}

@book{chen2016plasmaphysics,
author="Chen, Francis F.",

Title="Introduction to Plasma Physics and Controlled Fusion",
year="2016",
publisher="Springer International Publishing",

isbn="978-3-319-22309-4",
doi="10.1007/978-3-319-22309-4_1",}

@ARTICLE{DArcy2022,
      author       = {D'Arcy, R. and others},
      title        = {{R}ecovery time of a plasma-wakefield accelerator},
      journal      = {Nature},
      volume       = {603},
      issn         = {0028-0836},
      address      = {London [u.a.]},
      publisher    = {Nature Publ. Group},
      pages        = {58 - 62},
      year         = {2022},
      month        = {March},
      doi          = {10.1038/s41586-021-04348-8},
      url          = {https://bib-pubdb1.desy.de/record/459332},
}

@article{Decking2020xfel,
  title = {A MHz-repetition-rate hard X-ray free-electron laser driven by a superconducting linear accelerator},
  author = {Decking, W. and others},
  journal = {Nature Photonics},
  volume = {14},
  issue = {6},
  pages = {391–397},
  numpages = {6},
  year = {2007},
  month = {June},
  publisher = {Nature},
  doi = {10.1038/s41566-020-0607-z},
  url = {https://doi.org/10.1038/s41566-020-0607-z}
}

@article{doyle-2019,
author = {Doyle, B. L. and McDaniel, F. and Hamm, R. W.},
title = {The Future of Industrial Accelerators and Applications},
journal = {Reviews of Accelerator Science and Technology},
volume = {10},
number = {01},
pages = {93-116},
year = {2019},
month = {January}, 
doi = {10.1142/S1793626819300068},
}

@article{davidson2015osiris,
title = {Implementation of a hybrid particle code with a PIC description in r–z and a gridless description in $\Phi$ into OSIRIS},
journal = {Journal of Computational Physics},
volume = {281},
pages = {1063-1077},
year = {2015},
month = {January},
issn = {0021-9991},
doi = {10.1016/j.jcp.2014.10.064},
url = {https://www.sciencedirect.com/science/article/pii/S0021999114007529},
author = {A. Davidson and A. Tableman and W. An and F.S. Tsung and W. Lu and J. Vieira and R.A. Fonseca and L.O. Silva and W.B. Mori}
}

@article{downer2018diagnostic,
  title = {Diagnostics for plasma-based electron accelerators},
  author = {Downer, M. C. and Zgadzaj, R. and Debus, A. and Schramm, U. and Kaluza, M. C.},
  journal = {Reviews of Modern Physics},
  volume = {90},
  issue = {3},
  pages = {035002},
  numpages = {62},
  year = {2018},
  month = {August},
  publisher = {American Physical Society},
  doi = {10.1103/RevModPhys.90.035002},
  url = {https://link.aps.org/doi/10.1103/RevModPhys.90.035002}
}

@article{diederichs2023temperature,
    author = {Diederichs, S. and Benedetti, C. and Esarey, E. and Thévenet, M. and Sinn, A. and Osterhoff, J. and Schroeder, C. B.},
    title = {Temperature effects in plasma-based positron acceleration schemes using electron filaments},
    journal = {Physics of Plasmas},
    volume = {30},
    number = {7},
    pages = {073104},
    year = {2023},
    month = {07},
    issn = {1070-664X},
    doi = {10.1063/5.0155489},
    url = {https://doi.org/10.1063/5.0155489},
}

@article{esarey2009lasersrev,
  title = {Physics of laser-driven plasma-based electron accelerators},
  author = {Esarey, E. and Schroeder, C. B. and Leemans, W. P.},
  journal = {Reviews of Modern Physics},
  volume = {81},
  issue = {3},
  pages = {1229--1285},
  numpages = {0},
  year = {2009},
  month = {August},
  publisher = {American Physical Society},
  doi = {10.1103/RevModPhys.81.1229},
  url = {https://link.aps.org/doi/10.1103/RevModPhys.81.1229}
}

@article{ferrario2013sparclab,
title = "{SPARC}\_{LAB} present and future",
journal = {Nuclear Instruments and Methods in Physics Research Section B: Beam Interactions with Materials and Atoms},
volume = {309},
pages = {183-188},
year = {2013},
month = {August},
issn = {0168-583X},
doi = {10.1016/j.nimb.2013.03.049},
url = {https://www.sciencedirect.com/science/article/pii/S0168583X13003844},
author = {M. Ferrario and others}
}

@article{gorbunov2001plasma,
  title = {Plasma Ions Dynamics in the Wake of a Short Laser Pulse},
  author = {Gorbunov, L. M. and Mora, P. and Solodov, A. A.},
  journal = {Physical Review Letters},
  volume = {86},
  issue = {15},
  pages = {3332--3335},
  numpages = {0},
  year = {2001},
  month = {Apr},
  publisher = {American Physical Society},
  doi = {10.1103/PhysRevLett.86.3332},
  url = {https://link.aps.org/doi/10.1103/PhysRevLett.86.3332}
}

@article{Gilljohann2019ionmotion,
  title = {Direct Observation of Plasma Waves and Dynamics Induced by Laser-Accelerated Electron Beams},
  author = {Gilljohann, M. F. and others},
  journal = {Physical Review X},
  volume = {9},
  issue = {1},
  pages = {011046},
  numpages = {13},
  year = {2019},
  month = {March},
  publisher = {American Physical Society},
  doi = {10.1103/PhysRevX.9.011046},
  url = {https://link.aps.org/doi/10.1103/PhysRevX.9.011046}
}

@article{gonsalves2019initialplasmatemperature,
  title = {Petawatt Laser Guiding and Electron Beam Acceleration to 8 GeV in a Laser-Heated Capillary Discharge Waveguide},
  author = {Gonsalves, A. J. and others},
  journal = {Physical Review Letters},
  volume = {122},
  issue = {8},
  pages = {084801},
  numpages = {6},
  year = {2019},
  month = {Feb},
  publisher = {American Physical Society},
  doi = {10.1103/PhysRevLett.122.084801},
  url = {https://link.aps.org/doi/10.1103/PhysRevLett.122.084801}
}

@article{gorbunov2003ionmotion,
    author = {Gorbunov, L. M. and Mora, P. and Solodov, A. A.},
    title = {Dynamics of a plasma channel created by the wakefield of a short laser pulse},
    journal = {Physics of Plasmas},
    volume = {10},
    number = {4},
    pages = {1124-1134},
    year = {2003},
    month = {04},
    issn = {1070-664X},
    doi = {10.1063/1.1559011},
    url = {https://doi.org/10.1063/1.1559011},
}

@article{gholizadeh-2011,
  title     = {Effect of temperature on ion motion in future plasma wakefield accelerators},
  author    = {Gholizadeh, R. and Katsouleas, T. and Huang, C. and Mori, W. B. and Muggli, P.},
  journal   = {Physical Review Accelerators and Beams},
  volume    = {14},
  issue     = {2},
  pages     = {021303},
  numpages  = {5},
  year      = {2011},
  month     = {February},
  publisher = {American Physical Society},
  doi       = {10.1103/PhysRevSTAB.14.021303},
  url       = {https://link.aps.org/doi/10.1103/PhysRevSTAB.14.021303}
}

@article{huang2006quickpic,
title = {QUICKPIC: A highly efficient particle-in-cell code for modeling wakefield acceleration in plasmas},
journal = {Journal of Computational Physics},
volume = {217},
number = {2},
pages = {658-679},
year = {2006},
issn = {0021-9991},
doi = {https://doi.org/10.1016/j.jcp.2006.01.039},
url = {https://www.sciencedirect.com/science/article/pii/S0021999106000283},
author = {C. Huang and V.K. Decyk and C. Ren and M. Zhou and W. Lu and W.B. Mori and J.H. Cooley and T.M. Antonsen and T. Katsouleas},
}

@Article{hooker2013RepetitionRateNeeds,
author={Hooker, S. M.},
title={Developments in laser-driven plasma accelerators},
journal={Nature Photonics},
year={2013},
month={Oct},
day={01},
volume={7},
number={10},
pages={775-782},
issn={1749-4893},
doi={10.1038/nphoton.2013.234},
url={https://doi.org/10.1038/nphoton.2013.234}
}

@article{khudiakov-2022,
  doi       = {10.1088/1361-6587/ac4523},
  url       = {https://dx.doi.org/10.1088/1361-6587/ac4523},
  year      = {2022},
  month     = {February},
  publisher = {IOP Publishing},
  volume    = {64},
  number    = {4},
  pages     = {045003},
  author    = {V. K. Khudiakov and K. V. Lotov and M. C. Downer},
  title     = {Ion dynamics driven by a strongly nonlinear plasma wake},
  journal   = {Plasma Physics and Controlled Fusion}
}

@article{lehe2016fbpic,
  title     = {A spectral, quasi-cylindrical and dispersion-free Particle-In-Cell algorithm},
  journal   = {Computer Physics Communications},
  volume    = {203},
  pages     = {66-82},
  year      = {2016},
month       = {June},
  issn      = {0010-4655},
  doi       = {10.1016/j.cpc.2016.02.007},
  url       = {https://www.sciencedirect.com/science/article/pii/S0010465516300224},
  author    = {R. Lehe and M. Kirchen and I. A. Andriyash and B. B. Godfrey and J. Vay}
}

@misc{lindstrom2025pwfareview,
      title={Beam-driven plasma-wakefield acceleration}, 
      author={C. A. Lindstrøm and S. Corde and R. D'Arcy and S. Gessner and M. Gilljohann and M. J. Hogan and J. Osterhoff},
      year={2025},
      eprint={2504.05558},
      archivePrefix={arXiv},
      primaryClass={physics.acc-ph},
      url={https://arxiv.org/abs/2504.05558}, 
}

@article{Litos20169GeV,
doi = {10.1088/0741-3335/58/3/034017},
url = {https://dx.doi.org/10.1088/0741-3335/58/3/034017},
year = {2016},
month = {February},
publisher = {IOP Publishing},
volume = {58},
number = {3},
pages = {034017},
author = {M. Litos and others},
title = "9 {G}e{V} energy gain in a beam-driven plasma wakefield accelerator",
journal = {Plasma Physics and Controlled Fusion}
}

@article{lu-2006,
  title     = {Nonlinear Theory for Relativistic Plasma Wakefields in the Blowout Regime},
  author    = {Lu, W. and Huang, C. and Zhou, M. and Mori, W. B. and Katsouleas, T.},
  journal   = {Physical Review Letters},
  volume    = {96},
  issue     = {16},
  pages     = {165002},
  numpages  = {4},
  year      = {2006},
  month     = {April},
  publisher = {American Physical Society},
  doi       = {10.1103/PhysRevLett.96.165002},
  url       = {https://link.aps.org/doi/10.1103/PhysRevLett.96.165002}
}

@article{lu2010bubblesizescaling,
doi = {10.1088/1367-2630/12/8/085002},
url = {https://doi.org/10.1088/1367-2630/12/8/085002},
year = {2010},
month = {aug},
publisher = {},
volume = {12},
number = {8},
pages = {085002},
author = {Lu, W and An, W and Zhou, M and Joshi, C and Huang, C and Mori, W B},
title = {The optimum plasma density for plasma wakefield excitation in the blowout regime},
journal = {New Journal of Physics}
}

@article{massimo2016comparisons,
  title={Comparisons of time explicit hybrid kinetic-fluid code architect for plasma wakefield acceleration with a full pic code},
  author={Massimo, F. and Atzeni, S. and Marocchino, A.},
  journal={Journal of Computational Physics},
  volume={327},
  pages={841-850},
  year={2016},
  month={December},
  publisher={Elsevier},
 doi = {https://doi.org/10.1016/j.jcp.2016.09.067}
}

@book{macchi2013superintense,
  title     ={A superintense laser-plasma interaction theory primer},
  author    ={Macchi, A.},
  isbn      = {978-94-007-6124-7},
  doi       = {10.1007/978-94-007-6125-4},
  year      = {2013},
  month     = {January},
  publisher = {Springer Dordrecht}
}

@article{Mehrling2014hipace,
doi = {10.1088/0741-3335/56/8/084012},
url = {https://dx.doi.org/10.1088/0741-3335/56/8/084012},
year = {2014},
month = {July},
publisher = {IOP Publishing},
volume = {56},
number = {8},
pages = {084012},
author = {T. Mehrling and C. Benedetti and C. B. Schroeder and J. Osterhoff},
title = {HiPACE: a quasi-static particle-in-cell code},
journal = {Plasma Physics and Controlled Fusion}
}

@article{parise-2022,
    author  = {Parise, G. and Cianchi, A. and Del Dotto, A. and Guglietta, F. and Rossi, A. R. and Sbragaglia, M.},
    title   = {Lattice Boltzmann simulations of plasma wakefield acceleration},
    journal = {Physics of Plasmas},
    volume  = {29},
    number  = {4},
    pages = {043903},
    year    = {2022},
    month   = {April},
    issn    = {1070-664X},
    doi     = {10.1063/5.0085192},
    url     = {10.1063/5.0085192},
   _eprint  = {https://pubs.aip.org/aip/pop/article-pdf/doi/10.1063/5.0085192/16613003/043903\_1\_online.pdf},
}

@article{pompili2024recoverytime,
  title={Recovery of hydrogen plasma at the sub-nanosecond timescale in a plasma-wakefield accelerator},
  author={Pompili, R. and others},
  journal={Communications Physics},
  volume={7},
  number={1},
  pages={241},
  year={2024},
  month={July},
  publisher={Nature Publishing Group UK London},
  doi={10.1038/s42005-024-01739-x},
  url={https://www.nature.com/articles/s42005-024-01739-x}
}

@article{pompili2021enspreadmin,
  title={Energy spread minimization in a beam-driven plasma wakefield accelerator},
  author={R. Pompili and others},
  journal={Nature Physics},
  volume={17},
  issue={4},
  pages={499-503},
  year={2021},
  month={April},
  publisher={Springer Nature},
  doi={10.1038/s41567-020-01116-9},
  url={https://doi.org/10.1038/s41567-020-01116-9}
}

@article{pompili2022fel,
  title={Free-electron lasing with compact beam-driven plasma wakefield accelerator},
  author={R. Pompili and others},
  journal={Nature},
  volume={605},
  issue={7911},
  pages={659-662},
  year={2022},
  month={May},
  publisher={Springer Nature},
  doi={10.1038/s41586-022-04589-1},
  url={https://doi.org/10.1038/s41586-022-04589-1}
}

@article{rosenzweig2005effects,
  title = {Effects of Ion Motion in Intense Beam-Driven Plasma Wakefield Accelerators},
  author = {Rosenzweig, J. B. and Cook, A. M. and Scott, A. and Thompson, M. C. and Yoder, R. B.},
  journal = {Physical Review Letters},
  volume = {95},
  issue = {19},
  pages = {195002},
  numpages = {4},
  year = {2005},
  month = {Oct},
  publisher = {American Physical Society},
  doi = {10.1103/PhysRevLett.95.195002},
  url = {https://link.aps.org/doi/10.1103/PhysRevLett.95.195002}
}

@book{rezzolla-2013,
    author    = {Rezzolla, L. and Zanotti, O.},
    title     = {Relativistic Hydrodynamics},
    year      = {2013},
    isbn      = {978-0-19-880759-9},
    doi = {10.1093/acprof:oso/9780198528906.001.0001},
    publisher = {Oxford University Press}
}

@article{sahai2017ionmotion,
  title = {Excitation of a nonlinear plasma ion wake by intense energy sources with applications to the crunch-in regime},
  author = {Sahai, Aakash A.},
  journal = { Physical Review Accelerators and Beams},
  volume = {20},
  issue = {8},
  pages = {081004},
  numpages = {15},
  year = {2017},
  month = {Aug},
  publisher = {American Physical Society},
  doi = {10.1103/PhysRevAccelBeams.20.081004},
  url = {https://link.aps.org/doi/10.1103/PhysRevAccelBeams.20.081004}
}

@misc{sheehy-2024,
      title={Applications of Particle Accelerators}, 
      author={Suzie Sheehy},
      year={2024},
      eprint={2407.10216},
      archivePrefix={arXiv},
      primaryClass={physics.acc-ph},
      url={https://arxiv.org/abs/2407.10216}, 
}

@article{simeoni2024thermalLB,
    author = {Simeoni, Daniele and Parise, Gianmarco and Guglietta, Fabio and Rossi, Andrea Renato and Rosenzweig, James and Cianchi, Alessandro and Sbragaglia, Mauro},
    title = {Lattice Boltzmann method for warm fluid simulations of plasma wakefield acceleration},
    journal = {Physics of Plasmas},
    volume = {31},
    number = {1},
    pages = {013904},
    year = {2024},
    month = {January},
    issn = {1070-664X},
    doi = {10.1063/5.0175910},
    url = {https://doi.org/10.1063/5.0175910},
}

@misc{simeoni2025thermalwakestructure,
      title={Thermal wakefield structure in plasma acceleration processes: insights from fluid models and PIC simulations}, 
      author={Daniele Simeoni and Andrea Renato Rossi and Gianmarco Parise and Fabio Guglietta and Mauro Sbragaglia},
      year={2025},
      eprint={2512.20150},
      archivePrefix={arXiv},
      primaryClass={physics.plasm-ph},
      url={https://arxiv.org/abs/2512.20150}, 
}

@article{simeoni2024thermalfluid,
    author = {Simeoni, D. and Rossi, A. R. and Parise, G. and Guglietta, F. and Sbragaglia, M.},
    title = {Thermal fluid closures and pressure anisotropies in numerical simulations of plasma wakefield acceleration},
    journal = {Physics of Plasmas},
    volume = {31},
    number = {9},
    pages = {093103},
    year = {2024},
    month = {September},
    issn = {1070-664X},
    doi = {10.1063/5.0216707},
    url = {https://doi.org/10.1063/5.0216707},
}

@article{sosedkin2016lcode,
title = {LCODE: A parallel quasistatic code for computationally heavy problems of plasma wakefield acceleration},
journal = {Nuclear Instruments and Methods in Physics Research Section A: Accelerators, Spectrometers, Detectors and Associated Equipment},
volume = {829},
pages = {350-352},
year = {2016},
note = {$2^{nd}$ European Advanced Accelerator Concepts Workshop - EAAC 2015},
issn = {0168-9002},
doi = {https://doi.org/10.1016/j.nima.2015.12.032},
url = {https://www.sciencedirect.com/science/article/pii/S0168900215016034},
author = {A.P. Sosedkin and K.V. Lotov},
}

@article{shwab2013shadowgraphy,
    author = {Schwab, M. B. and others.},
    title = "{Few-cycle optical probe-pulse for investigation of relativistic laser-plasma interactions}",
    journal = {Applied Physics Letters},
    volume = {103},
    number = {19},
    pages = {191118},
    year = {2013},
    month = {November},
    issn = {0003-6951},
    doi = {10.1063/1.4829489},
    url = {10.1063/1.4829489},
}

@article{Tomassini2016qfluid,
doi = {10.1088/0741-3335/58/3/034001},
url = {https://dx.doi.org/10.1088/0741-3335/58/3/034001},
year = {2015},
month = {December},
publisher = {IOP Publishing},
volume = {58},
number = {3},
pages = {034001},
author = {Tomassini, P. and Rossi, A. R.},
title = {Matching strategies for a plasma booster},
journal = {Plasma Physics and Controlled Fusion}
}

@article{vieira2014ion,
    author = {Vieira, J. and Fonseca, R. A. and Mori, W. B. and Silva, L. O.},
    title = {Ion motion in the wake driven by long particle bunches in plasma},
    journal = {Physics of Plasmas},
    volume = {21},
    number = {5},
    pages = {056705},
    year = {2014},
    month = {05},
    issn = {1070-664X},
    doi = {10.1063/1.4876620},
    url = {https://doi.org/10.1063/1.4876620},
}

@article{vay2008pusher,
    author = {Vay, J.-L.},
    title = {Simulation of beams or plasmas crossing at relativistic velocitya)},
    journal = {Physics of Plasmas},
    volume = {15},
    number = {5},
    pages = {056701},
    year = {2008},
    month = {02},
    issn = {1070-664X},
    doi = {10.1063/1.2837054},
    url = {https://doi.org/10.1063/1.2837054},
}

@misc{vay2018WarpX,
  author = {Vay, Jean-Luc and others},
    title     = {{WarpX: An advanced Particle-In-Cell code}},
    year      = 2018,
    publisher = {Zenodo},
    doi       = {10.5281/zenodo.4571577},
    note      = {\url{https://doi.org/10.5281/zenodo.4571577}},
    howpublished = {https://blast-warpx.github.io}
 }

@article{vay2016picreviw,
author = {Vay, Jean-Luc and Lehe, R\'{e}mi},
title = {Simulations for Plasma and Laser Acceleration},
journal = {Reviews of Accelerator Science and Technology},
volume = {09},
number = {},
pages = {165-186},
year = {2016},
doi = {10.1142/S1793626816300085},
URL = {https://doi.org/10.1142/S1793626816300085},
}

@article{viera2012ionmotion,
  title = {Ion Motion in Self-Modulated Plasma Wakefield Accelerators},
  author = {Vieira, J. and Fonseca, R. A. and Mori, W. B. and Silva, L. O.},
  journal = { Physical Review Letters },
  volume = {109},
  issue = {14},
  pages = {145005},
  numpages = {5},
  year = {2012},
  month = {Oct},
  publisher = {American Physical Society},
  doi = {10.1103/PhysRevLett.109.145005},
  url = {https://link.aps.org/doi/10.1103/PhysRevLett.109.145005}
}

@article{yee-1966,
  author    = {K. Yee},
  journal   = {IEEE Transactions on Antennas and Propagation}, 
  title     = {Numerical solution of initial boundary value problems involving Maxwell's equations in isotropic media}, 
  year      = {1966},
  month     = {May},
  volume    = {14},
  number    = {3},
  pages     = {302-307},
  doi       = {10.1109/TAP.1966.1138693},
  url       = {10.1109/TAP.1966.1138693}    
}

@article{zgadzaj-2020,
  author  = {Zgadzaj, Rafal and others},
  title   = {Dissipation of electron-beam-driven plasma wakes},
  journal = {Nature Communications},
  year    = {2020},
  month   = {Sep},
  day     =  {21},
  volume  = {11},
  number  = {1},
  pages   = {4753},
  issn    = {2041-1723},
  doi     = {10.1038/s41467-020-18490-w},
  url     = {https://doi.org/10.1038/s41467-020-18490-w}
}

\end{document}